\documentclass{emulateapj}
\shorttitle{Superimposed AGN in clusters}
\shortauthors{Dotti \& Ruszkowski}

\def\gsim{\;\rlap{\lower 2.5pt
 \hbox{$\sim$}}\raise 1.5pt\hbox{$>$}\;}
\def\lsim{\;\rlap{\lower 2.5pt
   \hbox{$\sim$}}\raise 1.5pt\hbox{$<$}\;}

\newcommand{\kms}{\,{\rm km\,s^{-1}}}
\newcommand{\msun}{\,{\rm M_\odot}}
\newcommand{\lsun}{\,{\rm L_\odot}}
\newcommand{\beq}{\begin{equation}}
\newcommand{\eeq}{\end{equation}}
\def\lt{\mathrel{\rlap{\lower 3pt\hbox{$\sim$}}\raise 2.0pt\hbox{$<$}}}
\def\gt{\mathrel{\rlap{\lower 3pt\hbox{$\sim$}} \raise 2.0pt\hbox{$>$}}}
\def\lsim{\mathrel{\rlap{\lower 3pt\hbox{$\sim$}}\raise 2.0pt\hbox{$<$}}}
\def\gsim{\mathrel{\rlap{\lower 3pt\hbox{$\sim$}} \raise 2.0pt\hbox{$>$}}}
\begin{document}
\bibliographystyle{apj}

\title{AGN pairs: chance superpositions or black hole binaries?} 

\author{M. Dotti}
\affil{Max Planck Institute for Astrophysics, 
Karl-Schwarzschild-Str. 1, 85741 Garching, Germany; \\
mdotti@mpa-garching.mpg.de}
\author{M. Ruszkowski}
\affil{Department of Astronomy, University of Michigan, 500
Church Street, Ann Arbor, MI, USA; \\
e-mail: mateuszr@umich.edu}
\affil{The Michigan Center for Theoretical Physics, Ann Arbor, MI, 48109, USA}

\begin{abstract}

Several active galactic nuclei (AGN) with multiple sets of emission
lines separated by over $2000 \kms$ have been observed recently. These
have been interpreted as being due to massive black hole (MBH) recoil
following a black hole merger, MBH binaries, or chance superpositions
of AGN in galaxy clusters.  Moreover, a number of double-peaked AGN
with velocity offsets of $\sim$ a few $10^{2}\kms$ have also been
detected and interpreted as being due to the internal kinematics of
the narrow line regions or MBH binary systems. Here we reexamine the
superposition model.  Using the {\it Millennium Run} we estimate the
total number of detectable AGN pairs as a function of the emission
line offset. We show that AGN pairs with high velocity line
separations up to $\sim$$2000\kms$ are very likely to be chance
superpositions of two AGN in clusters of galaxies for reasonable
assumptions about the relative fraction of AGN. No superimposed AGN
pairs are predicted for velocity offsets in excess of $\sim 3000\kms$
as the required AGN fractions would violate observational constraints.
The high velocity AGN pair numbers predicted here are competitive with
those predicted from the models relying on MBH recoil or MBH binaries.
However, the model fails to account for the largest emission line
velocity offsets that require the presence of MBH binaries.\\

\end{abstract}


\section{Introduction}

Active galactic nuclei (AGN) with multiple sets of emission lines
(EL)\footnote{Unless stated otherwise in the text, we collectively refer to AGN
broad and narrow emission lines as EL for brevity.}
redshifted by a large factor ($> 2000 \kms$) have
been recently discovered (SDSS J092712.65+294344.0, Komossa, Zhou \&
Lu 2008; SDSS J153636.22+044127.0, Boroson \& Lauer 2009; SDSS
J105041.35+345631.3, Shields et al. 2009a).  The observation of these
sources have triggered a huge theoretical effort aimed at
understanding their physical nature.

Two models have been proposed as possible explanations for all these
AGN. The first one requires the presence of a massive black hole (MBH)
binary in the nucleus of each source. The orbital motion of the two
components of the binary can result in two sets of EL at two different
redshifts if the binary is not orbiting in the plane of the sky
(e.g., Begelman, Blandford \& Rees 1980). The two sets of ELs would be
blue-- and red--shifted with respect to the rest frame redshift of the
host galaxy.  The second model involves the presence of two distinct
galaxies that are interacting or simply superimposed within rich
galaxy clusters (GCs; hereafter ``superposition model''; Heckman et
al. 2009; Shields, Bonning \& Salviander 2009b; Wrobel \& Laor 2009;
Decarli et al.  2009a; Shields et al. 2009a).  Massive GC have been
advocated as a possible explanation for the presence of two
superimposed/interacting galaxies with high relative velocity along
the line of sight ($v_{\rm rel} \gsim 1000 \kms$).  Apart from its
simplicity, the superposition model has two main virtues:\\ \indent
$\bullet$ It can be easily falsified by a non-detection of an
overdensity of galaxies in the vicinity of AGN that show multiple sets
of ELs (see Decarli, Reynolds \& Dotti 2009b).\\ \indent $\bullet$ The
model can account for some of the spectroscopic binary AGN candidates
that have been recently identified in the DEEP2 Galaxy Redshift Survey
(Comerford et al. 2009) and in the Sloan Digital Sky Survey (Smith et
al. 2009; Liu et al. 2009; Wang et al. 2009; Xu \& Komossa 2009), that
exhibit two sets of ELs at redshifts that imply relative velocities of
few hundreds of $\kms$. Some of those objects can be superpositions in
less massive clusters or groups of galaxies, or superpositions in
massive GCs between galaxies that move predominantly in the direction
tangential to the line of sight.\\

In this Letter we present an estimate of the total number of
superimposed AGN in GCs at $0.1<z<0.7$, as a function of their
relative velocity, and discuss the implications that this estimate has
on the interpretation of AGN with multiple sets of ELs.\\

\section{Methodology and results}

Our work is based on the results of the {\it Millennium Run} 
(hereafter MR, Springel et al. 2005). We select from the Virgo-Millennium
Database\footnote{http://www.g-vo.org/Millennium} all the GCs in the
MR with masses $\gt 10^{13} \msun$
\footnote{Without the loss of generality we call every bound structure
  above this threshold a GC and we do not distinguish between small
  groups and rich clusters.}, and redshift between 0.1 and 0.7.  The
redshift interval has been chosen to coincide with that used by Smith
et al. (2009).  The upper limit of $z$ has been set to keep [OIII] in
a clean part of the spectrum, in order to allow a clear spectroscopic
detection of two superimposed AGN.  The choice of the minimum redshift
limit reduces the number of the low luminosity AGN that would have
otherwise been missed at larger redshifts.

We search for the distribution of galaxies in each GC in the online
galaxy catalog of De Lucia \& Blaizot (2007).  We select only those
galaxies that are luminous enough to be observed, i.e., those with
SDSS $u, g, r, i$, and $z$ band magnitudes lower then the SDSS
magnitude limits. In this sample, we consider only the galaxies that
host MBH more massive than $10^5 \msun$.

We select the galaxies that would be observed as superimposed, given
the positions of the two galaxies in the MR. Specifically, we
consider only superpositions between galaxies in the same GC. This
additional criterion ensures that our results provide a lower limit
to the number of pairs,
because we neglect (1) the superpositions between field galaxies, (2)
between a field galaxy and a galaxy in a GC, and (3) between two galaxies
in two different GCs. Two galaxies are considered superimposed if
their angular separation is lower than 3 arcsec, i.e., the angular
resolution of the SDSS spectra.

We then integrate over redshift, i.e., we sum the pairs
detected in all GCs in different redshift snapshots of the MR
renormalizing the number of pairs in each redshift interval 
by the ratio between the
physical volume of the Universe and the comoving volume simulated in the MR.

\begin{figure}
\begin{center}
\includegraphics[angle=0,width=8.5cm]{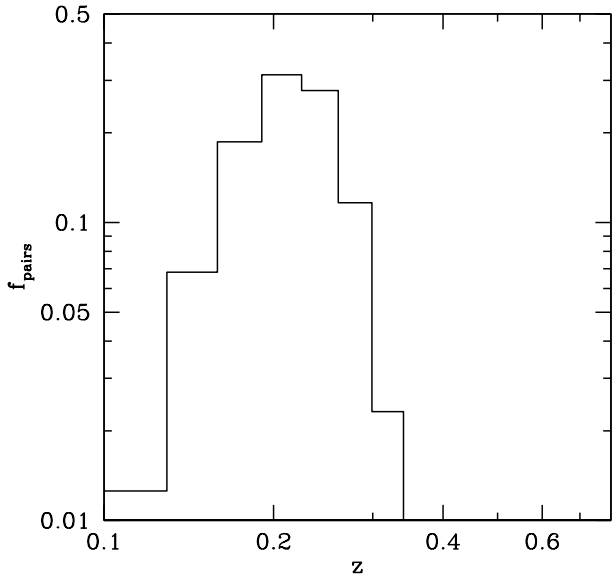}
\caption{Fraction of superimposed galaxy pairs as a function of
  redshift. The distribution of superimposed AGN is the same by
  construction.
\label{figz}}
\end{center}
\end{figure}

We renormalize the distribution of superimposed galaxies to compute
the distribution of superimposed AGN. Based on the complete fraction
of low redshift ($z<0.1$) SDSS galaxies, Kauffmann et al. (2004)
estimated that the fraction of AGN with [OIII] line luminosities
larger than $10^7 \lsun$ ($L_{\rm [OIII]}>10^7 \lsun$) is $f_{\rm AGN}
\sim 6 \%$ in high galaxy density regions.  Majority of galaxies in
the highest density regions (largest neighbor number) considered by
Kauffmann et al. (2004) reside in the dark matter halos of masses
exceeding $10^{13}\msun$, i.e., in galaxy groups and clusters (see
Figure 16, right panel in Kauffmann et al. 2004). The AGN fraction in
galaxies less massive then $\sim 10^{10} \msun$ is lower then the
assumed $f_{\rm AGN}$. However, the fraction of pairs with at least
one component less massive then $10^{10} \msun$ is typically around
$\sim$ 0.5\% and, thus, such low mass galaxies do not affect our
conclusions.  We limit our study to the case where both galaxies in a
given pair are active.  As a consequence, only $0.06^{2}$ of our pairs
are considered AGN pairs (i.e., 0.36\%). We note that, for a given
$f_{\rm AGN}$, the expected number of systems with multiple ELs
obtained under this assumption is a lower limit, as under particular
conditions two sets of ELs can be produced by an AGN/inactive galaxy
pair (Heckman et al. 2009). That is, a second set of narrow ELs can be
produced by an inactive galaxy if it is close enough to an AGN, so
that the gas is photoionized by the AGN continuum.  Our requirement
that both the galaxies are active assures that they can be observed as
multiple line emitters, regardless of their separation or their
luminosity.

Because our study concerns AGN at redshifts higher then those in the Kauffmann
et al. (2004) sample, it could in principle select a higher luminosity
subsample of them. The fraction of AGN is a decreasing function of the [OIII]
luminosity (see right panel of Figure 9 in Kauffmann et al. 2004), so at
larger redshift we expect lower $f_{\rm AGN}$.  Because our predicted
distribution of AGN pairs peaks at $z\approx 0.2$, the luminosity distances of
our systems are $\gsim 4$ times those of the AGN in Kauffmann sample. As a
consequence we could be selecting AGN with $L_{\rm [OIII]}>10^8 \lsun$, that
would correspond to $f_{\rm AGN}$ smaller by $\sim$ one order of
magnitude. Therefore, we also consider ten times lower $f_{\rm AGN}$ in order
to obtain a conservative lower number of superimposed AGN. However, a
significant fraction of AGN with $L_{\rm [OIII]}>10^7 \lsun$ AGN may still be
detectable at $z\sim 0.2$.  We also that the number of AGN in this sub-sample
is conservative also because, if AGN itself is luminous enough, it can be
observed even if the luminosity of the galaxy is too faint to be
detected. Finally, for low $z$ the intrinsic AGN fraction increases with
redshift.

Figure~\ref{figz} shows the fraction of superimposed galaxies (f$_{\rm
  pairs}$) as a function of redshift. We note that this distribution
is by construction equal to the distribution of superimposed AGN,
given that we assumed the AGN-to-galaxy ratio to be independent of
redshift. We note that this assumption does not strongly affect our
results because the distribution of galaxy pairs is strongly peaked at
$z \approx 0.2$ and does not extend to $z\gsim 0.4$.  The distribution
of galaxy pairs has a clear peak.  This distribution can be used to
used to constrain the models. For example, the AGN pairs at $z\ga 0.5$
are unlikely to be consistent with superposition model given the SDSS
limits. The peak in the distribution occurs for two reasons:
\\ \indent $\bullet$ The small number of pairs at low $z$ is due to
the fact that the volume of the universe enclosed within $z \lsim 0.1$
is relatively small and so is the number of GCs in that region.
Furthermore, a galaxy pair at a given physical separation may appear
to be separated by 3 arcsec and count as superimposed pair when it is
at higher redshift but may be resolved when it is at low $z$ thereby
lowering the expected superimposed pair fraction at small
redshifts.\\ \indent $\bullet$ At higher $z$, the number of massive
clusters decreases with redshift in accordance with hierarchical model
of structure formation.  The decline in the number of pairs at high z
is also caused by the SDSS magnitude limits.

\begin{figure}
\begin{center}
\includegraphics[angle=0,width=8.5cm]{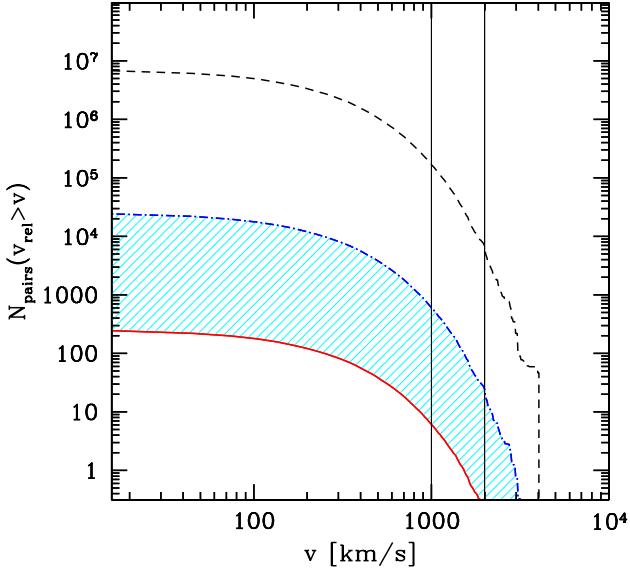}
\caption{Cumulative number of superimposed galaxy pairs (black dashed
  line) and AGN pairs (blue dot--dashed and red solid lines) with
  relative velocity along the line of sight larger than $v$. Blue
  (red) line has been obtained using $f_{\rm AGN}=6\% \, (0.6\%)$. The
  thin vertical lines highlight the number of pairs with relative
  velocity larger than 1000 and 2000 $\kms$, respectively.
\label{fig1}}
\end{center}
\end{figure}

Figure~\ref{fig1} shows the main results of our study.  The black
dashed line refers to the superimposed galaxy pairs. The blue
dot--dashed line refers to the AGN pairs, assuming $f_{\rm
  AGN}=6\%$. We find that over $\sim 10^4$ superimposed AGN are
expected from our simple model. The majority of these would not be
recognized as pairs because their relative velocity is too small.  The
spectra of superimposed pairs with $v_{\rm rel}\sim 100 \kms$ would
not appear peculiar.  In fact, the coexistence of two different sets
of ELs with correspondingly small redshift difference can be due to
the dynamics within a single galaxy (e.g., due to a disk-like
structure in the narrow line region or due to the presence of winds,
see, e.g., Crenshaw et al. 2009 and references therein). If two
superimposed AGN are Type I, and have such a small relative velocity,
it would be extremely difficult (or impossible) to deconvolve the two
broad line sets from the spectrum. Nevertheless, a significant number
of superimposed AGN with high relative velocities is present.  We
predict $N_{\rm pair}\gsim 500$ AGN pairs have $v_{\rm rel}>1000 \kms$,
and $N_{\rm pair}\approx 18$ pairs with $v_{\rm rel}>2000 \kms$.

These results seem to be in contrast with the paucity of observed AGN with
multiple sets of ELs (Comeford et al. 2009; Smith et al. 2009, Liu et
al. 2009; Wang et al. 2009; Xu et al. 2009). In these studies, the authors
find few hundred systems with a difference in redshift corresponding to
$v_{\rm rel}\gsim 100 \kms$, while we find $\approx 4000$ pairs with $v_{\rm
rel}\gsim 500 \kms$.  However, our results can match these observations if we
conservatively assume $f_{\rm AGN}=0.6\%$ (red solid line in
Figure~\ref{fig1}). In this case, the total number of AGN superpositions is
$\sim 200$.  Therefore, even for this conservative assumption for $f_{\rm
AGN}$, a considerable fraction of the binary AGN candidates with low relative
velocities can be explained as AGN superpositions in GCs. We emphasize that
AGN fraction higher then $0.6\%$ can be consistent with the data as the
surveys performed to date do not cover the whole sky.

As a consequence of the above change in the normalization in $f_{\rm
  AGN}$, the number of superpositions with $v_{\rm rel}>1000 \kms$ is
5, while no AGN pairs are found for $v_{\rm rel}>2000
\kms$. Therefore, in this conservative limit, none of the high
velocity superposition candidates (SDSS J092712.65+294344.0, SDSS
J153636.22+044127.0, and SDSS J105041.35+345631.3) would be consistent
with a superposition of two AGN in galaxy clusters.  However, the
superposition model could still account for the faint AGN-quasar pair
in SDSS J153636.22+044127.0 (see below).

We expect the real distribution of superimposed AGN in clusters to lie
in the region between the blue dot--dashed and the red solid lines in
Figure~\ref{fig1}. We note that for the case where either every galaxy in GCs hosts an AGN or,
equivalently, we do not require the two galaxies host AGN in order to
be spectroscopically detected as a pair ($f_{\rm AGN}=1$), we do not expect any pair with $v_{\rm rel}>4000 \kms$.

\section{Discussion}

We analyzed the dynamical properties of galaxies in clusters and groups of
galaxies in the De Lucia \& Blaizot galaxy catalog based on the {\it
Millennium Run}. We computed the total number of superimposed AGN pairs in the
whole sky as a function of galaxy relative velocity along the line of sight
and under the assumption that a fixed fraction of galaxies are in the active
phase and taking into account the SDSS magnitude and angular resolution
limits.  We find that, for $0.6 \% \lsim f_{\rm AGN} \lsim 6 \%$, the expected
cumulative number of AGN pairs with small velocity separations of $v_{\rm
rel}$ ($\gsim 100 \kms$) is always $\gsim 200$ within $z\approx 0.3$. In the
most optimistic case, this number increases to $\sim 2\times 10^4$.  This can
explain a substantial fraction (if not all) of the AGN with multiple sets of
ELs extracted from the SDSS and DEEP2 surveys (e.g., Liu et
al. 2009). However, some of those objects can be explained in terms of
internal dynamics of gas in the NLR (e.g., Crenshaw et al. 2009).

The AGN pairs with velocity separations $v_{\rm rel}\gsim 500 \kms$
are more difficult to explain in terms of internal dynamics of
single galaxies then those characterized by smaller velocity
differences. For $v_{\rm rel}\gsim 500\kms$, we find between $\sim 40$
and a few $\times 10^{3}$ AGN pairs depending on the assumed AGN
fraction which can account for a sizable fraction of the
observed AGN pairs at this velocity separation.

For $v_{\rm rel}\gsim 1000 \kms$, we find more than 5 AGN pairs. This
number can increase up to 500 if we optimistically assume that a large
fraction of galaxies in GCs host AGN ($f_{\rm AGN} = 6 \%$).  AGN
showing multiple sets of ELs corresponding to such extreme velocities
cannot be explained in terms of internal dynamics of a single galaxy
but can be easily accounted for in the superposition model.  We also
note that similar (or even larger) relative velocities are predicted
by two other models: (1) the binary model, and (2) the recoiling MBH
model where a set of ELs is comoving with a recoiling MBH -- a remnant
of a MBH binary coalescence (Komossa et al.  2008). However, even
these models predict very few events ($\lsim 20$ for the binary model,
Volonteri, Miller, Dotti 2009; $\lsim 1$ for the recoil model, Dotti
et al. 2009; Shields et al. 2009). Therefore, the superposition model
appears at least equally competitive in this case.

However, three AGN with multiple sets of emission lines shifted by
even larger amount (more than $2000 \kms$) have been recently
detected. It is interesting to compare our results with such extreme
sources.

SDSS J092712.65+294344.0 has two sets of ELs corresponding to
$v_{\rm}\approx 2650 \kms$.  From the theoretical point of view, it
could in principle be due to AGN superposition in a very massive GC if
we assume $f_{\rm AGN}\sim 6\%$. However, the redshift of the cluster
($z\approx 0.7$) is inconsistent with the redshift distribution of
superimposed AGN pairs that we obtained in this analysis (see
Figure~1).  Moreover, Decarli et al. (2009b) compared multiple band
images of the field containing SDSS J092712.65+294344.0 and proved
that the presence of such a massive GC is in disagreement with the
paucity of galaxies in that field.

SDSS J105041.35+345631.3 and SDSS J153636.22+044127.0 have multiple
sets of ELs with $v_{\rm rel}=3500 \kms$. Such high velocities are
inconsistent with any reasonable value of $f_{\rm AGN}$.  Because
these quasars show two distinct sets of broad lines, a model with a
recoiling MBH does not apply. For these objects other explanations,
implying the presence of MBH binaries or a non axisymmetric broad line
region (such as double peaked quasars see, e.g., Eracleous \& Halpern
2003; Gezari, Halpern \& Eracleous 2007), are required.

SDSS J153636.22+044127.0 has been the subject of follow--up
observations. Decarli et al. (2009a) found a significant overdensity
of galaxies that suggested the presence of a moderately rich cluster.
Furthermore, they discovered a second galaxy, within $\approx 1$
arcsec from the quasar host. The companion hosts an optically faint
AGN, clearly detected in radio by Wrobel et al. (2009). No direct
estimate of the redshift of the companion has been reported to date.
This quasar-AGN pair can be easily understood in terms of our model,
and could be an example of a case where one of the AGN is not bright
enough to possess clearly detectable emission lines.

Our investigation shows that an AGN pair with two sets of ELs shifted
by $500 \lsim v_{\rm rel} \lsim 2000 \kms$ is likely to be a chance
superposition of two AGN provided that the pair is located in dense
environment. On the other hand, the superposition model does not
predict sources with $v_{\rm rel}\gsim 3000 \kms$, even in the best
case scenario.  Even assuming that every galaxy is an AGN, no pairs
with $v_{\rm}>4000 \kms$ are expected to be observed.  The maximum
velocity for the superposition model is very similar to the maximum
recoil velocity predicted by fully relativistic simulations of MBH
coalescence ($v_{\rm rel} \approx 4000 \kms$, Campanelli et
al. 2007). AGN with larger shifts and multiple sets of narrow ELs
cannot be explained as standard double peaked emitters.  Such AGN are
likely to be close MBH binaries.

\acknowledgements

\end{document}